# On the discovery of K I 7699Å line strength variation during the 1982 – 1984 eclipse of ε Aurigae

M. Parthasarathy
Indian Institute of Astrophysics,
Koramangala, Bangalore 560 034
India

Abstract: The discovery of K I 7699Å line strength variations during the 1982-1984 eclipse of ε Aurigae is described. The equivalent widths and radial velocities of the K I 7699Å line derived from spectra obtained during 1981 November to 1983 July with the 2.1-meter Otto Struve reflector telescope of the McDonald observatory are presented.

1. Introduction

   ε Aurigae (HR 1605 = HD 31964) is a very long period (P=27. 1 years) eclipsing binary system consisting of an F0Iae supergiant and an invisible companion. The primary eclipse is total with a depth of magnitude 0.8 and the duration of the totality phase is 330 days. The secondary component of ε Aurigae has been an enigma since the eclipse of 1928. The eclipse is peculiar because the spectrum during the totality phase is the same as that of the primary component (F0Iae). The eclipse depth is independent of the wave length. There is no secondary minimum in the light curve. It is a single-lined spectroscopic binary.

   After the 1955 eclipse the earlier models (Kuiper et al. 1937, Struve 1956, Struve and Zebergs 1962) of the secondary were replaced by three or four models (Kopal 1954, 1971, Huang 1965, Wright 1970, Gyldenkerne 1970, Hack 1961, Cameron 1971, Wilson 1971).

   The proposed models of the secondary companion range from a swarm of meteorites to a black hole (Ludendorff 1924, Cameron 1971). Huang (1965) proposed that the secondary companion is an opaque disc of cool material, seen edge on.

   The spectroscopic and photometric results of the 1955 eclipse were discussed by Wright (1970), Gyldenkerne 1970, and Sahade and Wood (1978).

   The 1982-1984 eclipse of ε Aurigae was the first to have been extensively studied in the wavelength range from far-UV to the far-IR.

   The K I 7699Å line strength variation and hence the presence of neutral gas in and around the disc- shaped secondary of ε Aurigae was for the first time discovered by Parthasarathy (1982: see Stencel 1982: ε Aurigae News Letter No. 5, 5 (R.E Stencil 1982), Parthasarathy & Lambert (1983 a, b, c) from the systematic increase in the strength of K I 7699Å absorption line during the 1982 – 1984 eclipse. Hack &

Selvelli (1979) first observed the UV excess in the IUE UV spectrum of ε Aurigae. The IUE UV spectra obtained during the 1982-1984 eclipse also revealed the UV excess and indicated the presence of a B5V star at the center of the disc (Parathasarathy & Lambert 1983d, Chapman, Kondo & Stencel 1983, Ake & Simon 1983 & 1984, Boehm et al. 1984, Altner et al. 1986).

Light and color variations during the eclipse of ε Aurigae were discussed by Parthasarathy & Frueh (1986). The photometric contact dates of the 1982-1984 eclipse were derived from the light curve by Schmidtke 1985) and Parthasarathy & Frueh (1986).

Table 1 - Photometric contact dates of the eclipse of ε Aurigae during the 1982-1984 eclipse

$1^{st}$ contact HJD 2445165
$2^{nd}$ contact HJD 2445302
$3^{rd}$ contact HJD 2445748
$4^{th}$ contact HJD 2445812

2. Observations

I have obtained the spectra of ε Aurigae centered around the K I 7699Å line from 1981 November $11^{th}$ to 1983 July $19^{th}$ covering the pre-eclipse, ingress and totality phases of the eclipse with the 2.1m McDonald Observatory Otto Struve reflector telescope equipped with a Coude spectrograph and a Reticon detector.

The K I 7665A line is very much blended with the strong telluric O2 Lines. And hence it was not considered even though it is present in the observed spectra.

In this paper I report the K I 7699Å line equivalent widths and radial velocities. Though late, these results are important, to compare the K I 7699Å line strength variation with future eclipse observations of ε Aurigae, as well as for historians as a record of the discovery of the K I 7699Å line strength variation during the 1982 - 1984 eclipse of ε Aurigae.

(I had prepared in 1982/83 a first order draft of a paper intended as a letter for publication in NATURE journal in which I included the K I 7699Å line strength variation and results of IUE observations outside the eclipse to ingress to totality and gave it to David Lambert. He deleted the K I 7699Å line strength variation and gave me a first order draft intended for publication in the PASP which included only the results of the IUE observations (Parthasarathy and Lambert 1983d). Much before I left Austin (I left Austin in very early October 1983), I gave David Lambert the K I 7699Å line equivalent widths and radial velocities based my own observations from November 1981 to July 1983 and also those based on the observations made with the 2.7-m telescope from March 1982 to August 1983 (Lambert & Sawyer 1986), but these were never published.)

3. Analysis

The dates of observations, K I 7699Å line, equivalent widths and radial velocities are given in Table 2.

Table 2 – K I 7699Å line strength variation during the 1982 -1984 eclipse of ε Aurigae from the observations made with 2.1m Otto Struve telescope.

| Date | HJD 2440000+ | EW(mÅ) | R V (km/s) |
|---|---|---|---|
| 1981 Nov 11.315 | 4919.5 | 102 | +3 |
| 1981 Dec 20.168 | 4958.5 | 110 | +3 |
| 1982 April 03.151 | 5062.5 | 120 | +4 |
| 1982 July 27.461 | 5177.5 | 270 | +13 |
| 1982 Oct 01.390 | 5243.5 | 440 | +21 |
| 1982 Nov 02.257 | 5275.5 | 500 | +23 |
| 1982 Nov 04.523 | 5277.5 | 480 | +21 |
| 1982 Dec 28.094 | 5331.5 | 550 | +21 |
| 1983 Feb 27.201 | 5393.5 | 580 | +16 |
| 1983 Mar 01.143 | 5394.5 | 540 | +12 |
| 1983 July 19th | 5535.1 | 440 | -6.0 |

I have also observed and coordinated the K I 7699Å line observations and entire observations of the spectrum of ε Aurigae with the McDonald Observatory 2.7-m reflector telescope and Coude spectrometer (Tull et al. 1975, Vogt et al. 1978) from 1982 March to 1983 September and I have derived the equivalent widths and radial velocities of K I 7699Å line including my own observations from 1981 November to 1983 July with the 2.1-m Otto Struve reflector. I have included a figure of the K I 7699Å line variation in the IUE observing proposal that I prepared. It is shown in Figure 1 in this paper.

The 1981 Nov 11th spectrum is outside the eclipse and the eclipse by the neutral gas of the disc had not started then. The 1981 Nov 11th K I 7699Å line strength of 102mA is of interstellar origin (Table 2). This is in agreement with the strength of the interstellar component derived from the very high resolution (75000) spectra (104mA±1, Muthumariappan et al. 2014). In an F0Ia star, the photospheric lines of K I 7699Å & KI 7665 are not expected as the temperature of the star is relatively higher for K to be in the neutral state. The eclipse by the neutral gas of the disc- shaped secondary started about 90 days in advance of first contact of the photometric eclipse (Table 1). The K I line strength and radial velocity variation observed during the 1982 – 1984 eclipse repeats consistently during the 2009 – 2011 eclipse (Leadbeater et al. 2012, Potravnov & Grinin 2013, and Muthumariappan et al. 2014). Based on the study of spectra of previous eclipses Griffin & Stencel (2013) concluded that the structure of the disc does not alter appreciably on time scale of a century and they discovered a mass transfer stream from the F0Ia star on to the disc. Recently Stencel et al. (2015) find transient CO absorption and persistent Brackett Alpha emission and $^{12}C/^{13}C$ = 5 in the disc.

The interferometric images published by Kloppenborg et al. (2010 & 2015) clearly revealed that the eclipsing object is an asymmetric disc and derived the characteristics of the disc.

Muthumariappan & Parthasarathy (2012), and Muthumariappan et al. (2014) presented the structure of the disc- shaped secondary and post-AGB evolutionary status of the F0Ia primary of Epsilon Aurigae from 2D radiative transfer modelling and from the K I 7699Å line data respectively. A comprehensive summary of the results from the analysis of data obtained during the eclipses can be found in the paper by Stencel (2012).

Budaj (2011) studied the effects of dust on light curves of ε Aurigae type binary stars. Budaj's (2011) models of ε Aurigae consist of two geometrically thick flared discs: an internal optically thick disc and an external optically thin disc, which absorbs and scatters radiation. The discs are in the orbital plane and are almost edge – on. Budaj (2011) argues that there is no need for an inclined disc with a hole to explain the eclipse of ε Aurigae. He concludes that the shallow mid-eclipse brightening might result from the eclipse by nearly edge on flared (dusty and gaseous) discs.

Budaj (2013) modeled the light curve of KIC 12557548b, an extra- solar planet with a comet like tail. Budaj (2013) concludes that this light curve with pre-transit brightening is analogue to the light curve of ε Aurigae with mid-eclipse brightening and that forward scattering plays a significant role in such eclipsing systems.

Recent discoveries indicate that the very long period binaries TYC2505-672-1 (Lipunov et al. 2016, Rodriguez et al. 2016), OGLE-BLG182.1.162852, OGLE-LMC-ECL11893 and ELHC 10 in LMC. (Rattenbury et al. 2015, Garridio et al. 2016, Lipunov et al. 2016, Dong et al. 2014, and Meng et al. 2014) have disc- shaped secondary components which are similar to the disc- shaped secondary of ε Aurigae. Monitoring the K I 7699Å line equivalent widths and radial velocities during the eclipses will enable us to understand the origin and evolutionary stage of these systems and also that of Epsilon Aurigae.

Acknowledgements:-

I am very much indebted to late Prof. Harlan Smith for his kind encouragement, support, and for generously allotting observing time on the 2.1-meter Otto Struve reflector telescope of the McDonald observatory.

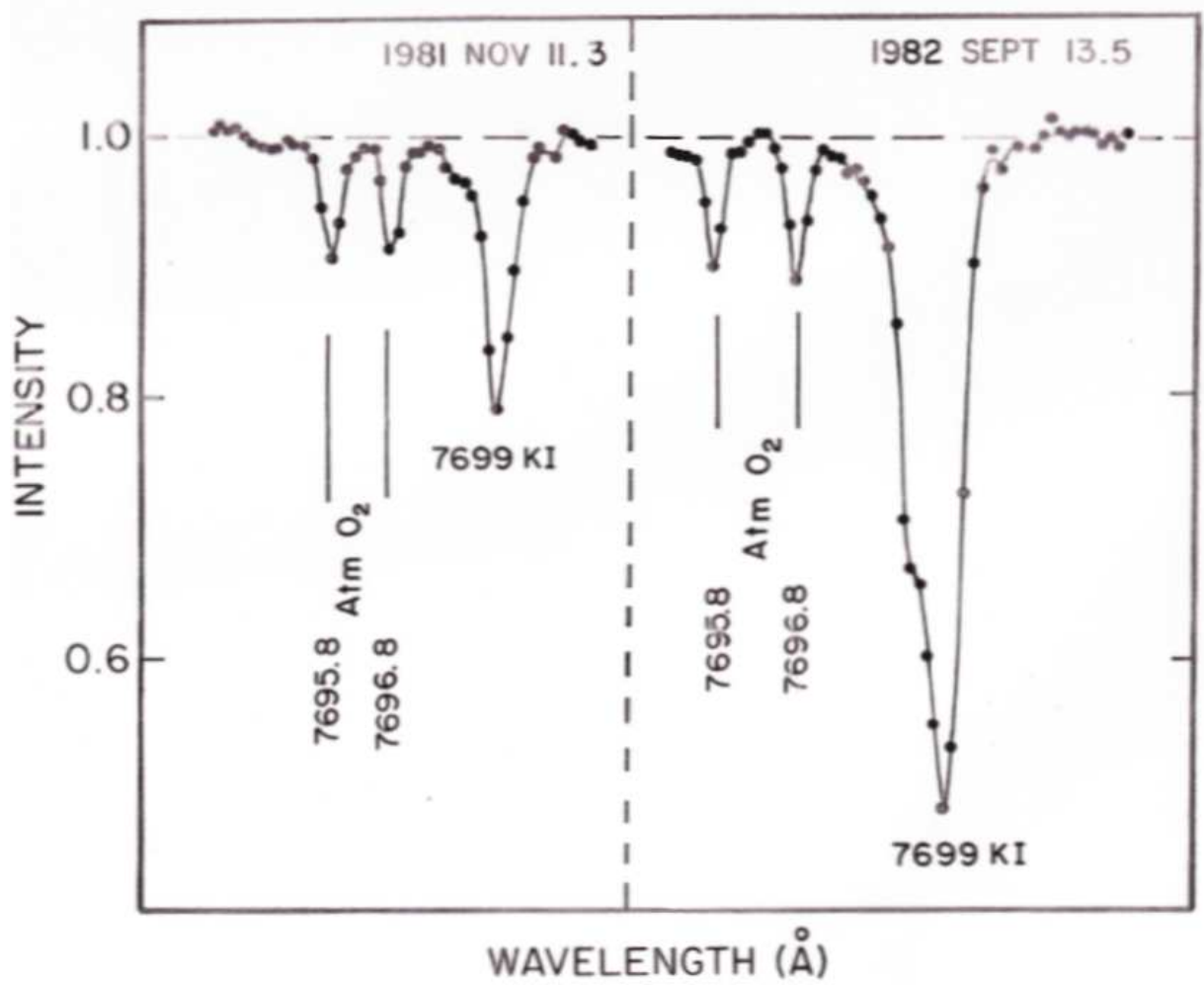

Figure 1. Example of K I 7699Å line strength variation during the eclipse of ε Aurigae